\documentclass[CEJP,DVI]{cej} 
\usepackage{graphicx}
\usepackage{layout}
\usepackage{amsmath}
\usepackage{textcomp}
\usepackage{hyperref}

\title{Centrality dependence of freeze-out parameters from Au+Au
  collisions at $\sqrt{s_{NN}}=$7.7, 11.5 and 39 GeV}

\articletype{Research Article}

\author{Lokesh Kumar\email{lokesh@rcf.rhic.bnl.gov}(for the STAR collaboration)}

\institute{Kent State University, Kent, OH-44242, USA}

\abstract{The RHIC beam energy scan program in its first phase collected
 data for Au+Au collisions at beam energies of 7.7, 11.5 and 
 39 GeV. The event statistics collected at these lower energies 
 allow us to study the centrality dependence of various observables
 in detail, and compare to fixed-target experiments at SPS for
 similar beam energies. The chemical and kinetic freeze-out
 parameters can be extracted from the experimentally measured
 yields of identified hadrons within the framework of thermodynamical
 models. These then provide information about the system at the
 stages of the expansion where inelastic and elastic collisions of
 the constituents cease. 
 We present the centrality dependence of freeze-out parameters
 for Au+Au collisions at midrapidity for $\sqrt{s_{NN}} = 7.7$,
 11.5, and 39 GeV from the STAR experiment. The chemical freeze-out
 conditions are obtained by comparing the measured particle ratios
 (involving $\pi$, $K$, $p$, and $\bar{p}$)
 to those from the statistical thermal model calculations.
 The kinetic freeze-out conditions are extracted at these energies
 by simultaneously fitting the invariant yields of identified
 hadrons ($\pi$, $K$, and $p$) using Blast Wave model calculations.
}


\begin{document}
\maketitle



\section{Introduction}

The experiments at the Relativistic Heavy-Ion Collider (RHIC) aim to
look for the signatures of the production of a Quark-Gluon Plasma (QGP). The
RHIC Beam Energy Scan (BES)~\cite{bes} program is devoted to study the QCD phase
diagram which involves searching for the QCD phase boundary and the QCD
critical point~\cite{qcdcp}. 
The QCD phase diagram is usually represented by plotting temperature $T$
vs. baryon chemical potential $\mu_{\rm{B}}$.
There are two main phases expected in
the QCD phase diagram - QGP and hadronic gas phase. 
The plan of the RHIC BES is to collect data at different
center-of-mass energies and look for the possible signatures of a QGP
and QCD critical point. 
It might then also be possible to locate the beam energy below which
there is no QGP formation.
To access the phase diagram, one needs 
$T$ and $\mu_{\rm{B}}$ corresponding to the beam energy, which can be obtained
from the spectra and ratios of produced particles. 

The bulk properties of particle production in heavy-ion collisions can be
studied using identified particle spectra. The measurements of
particle
abundances and transverse momentum distributions could provide
information about the final stages of the collision evolution at chemical
and kinetic freeze-out. 
The statistical thermal model~\cite{stm} has successfully described the ratios of
hadron yields produced in heavy-ion collisions from lower to higher
energies. 
The centrality dependence of model parameters, especially at low
energies could provide further understanding of the collision
dynamics. 


In this paper, we present the first measurements
of centrality dependence of freeze-out parameters at
$\sqrt{s_{NN}}=$7.7, 11.5 and 39 GeV. The results are 
presented for midrapidity $|y|<0.1$ region.
The data are taken by the STAR experiment for Au+Au
collisions in the year 2010. 
The total events analyzed are about 4 M, 8 M, and 10 M, respectively for $\sqrt{s_{NN}}$ = 7.7, 11.5, and 39 GeV.
The centrality selection is done using the uncorrected charged track
multiplicity
measured event-wise in the TPC within $|\eta|<$ 0.5. Centrality classes represent the
fractions of this multiplicity distribution.  
The average numbers of 
participating nucleons $\langle N_{\rm{part}} \rangle$ 
are obtained
by comparing the multiplicity distribution with that from Glauber
Monte simulation~\cite{bes}.



\section{Results and Discussions}
The transverse momentum spectra of hadrons ($\pi$, $K$, and
$p$) are obtained for the Au+Au collisions at $\sqrt{s_{NN}}=$7.7, 11.5
and 39 GeV and different centralities~\cite{lok_qm2011}. 
The proton spectra are not feed-down corrected for weak decays.
The particle yields and ratios are used to obtain the
freeze-out conditions. 
The statistical thermal model
(THERMUS)~\cite{stm} is used to extract the chemical freeze-out
conditions at these energies.
Kinetic freeze-out conditions are obtained
by fitting the spectra with the Blast-Wave (BW) model
calculations~\cite{bw}. 

\subsection{Chemical freeze-out}
The scenario after the collisions when the inelastic collisions among
the particles stop is called the chemical freeze-out. At this stage,
the particle yields and ratios are finalized and do not change
afterwards. The system can be described by the thermal equilibrium
model (THERMUS). 
The model assumes thermal and chemical equilibrium. 
The main fit parameters are chemical freeze-out parameter
$T_{\rm{ch}}$, baryonic chemical potential $\mu_{\rm{B}}$, strangeness
chemical potential $\mu_{\rm{S}}$, and strangeness suppression factor
$\gamma_{s}$.
The grand-canonical (GC) approach is used to fit the experimental particle ratios
($\pi^{-}/\pi^{+}$, $K^{-}/K^{+}$, $\bar{p}/p$, 
$K^{-}/\pi^{-}$, and $\bar{p}/\pi^{-}$).
For GC ensemble, the baryon number, charge number, and
strangeness content are conserved on average. The model explains the data very well.

\begin{figure}
\includegraphics[scale=0.35]{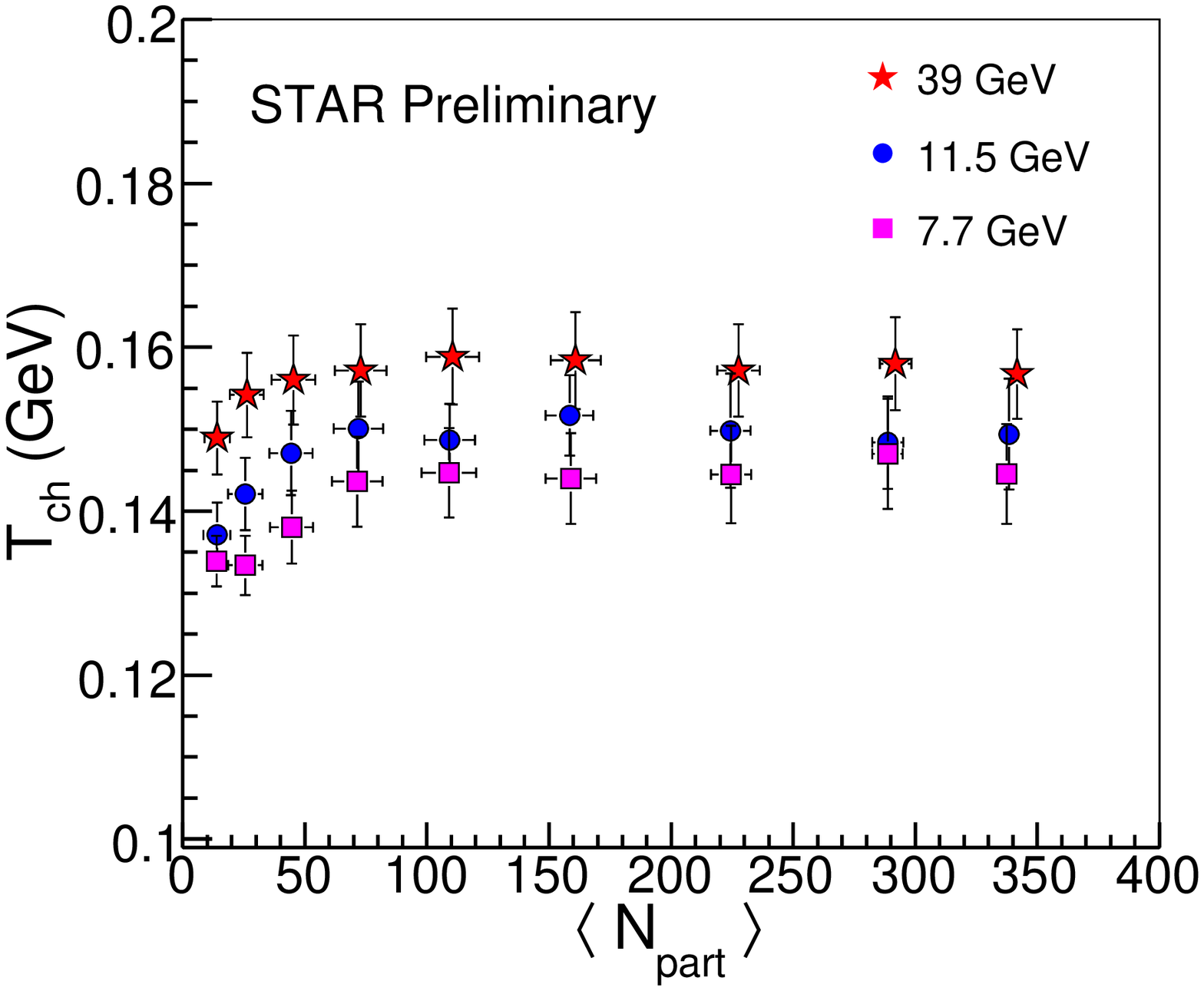}
\includegraphics[scale=0.35]{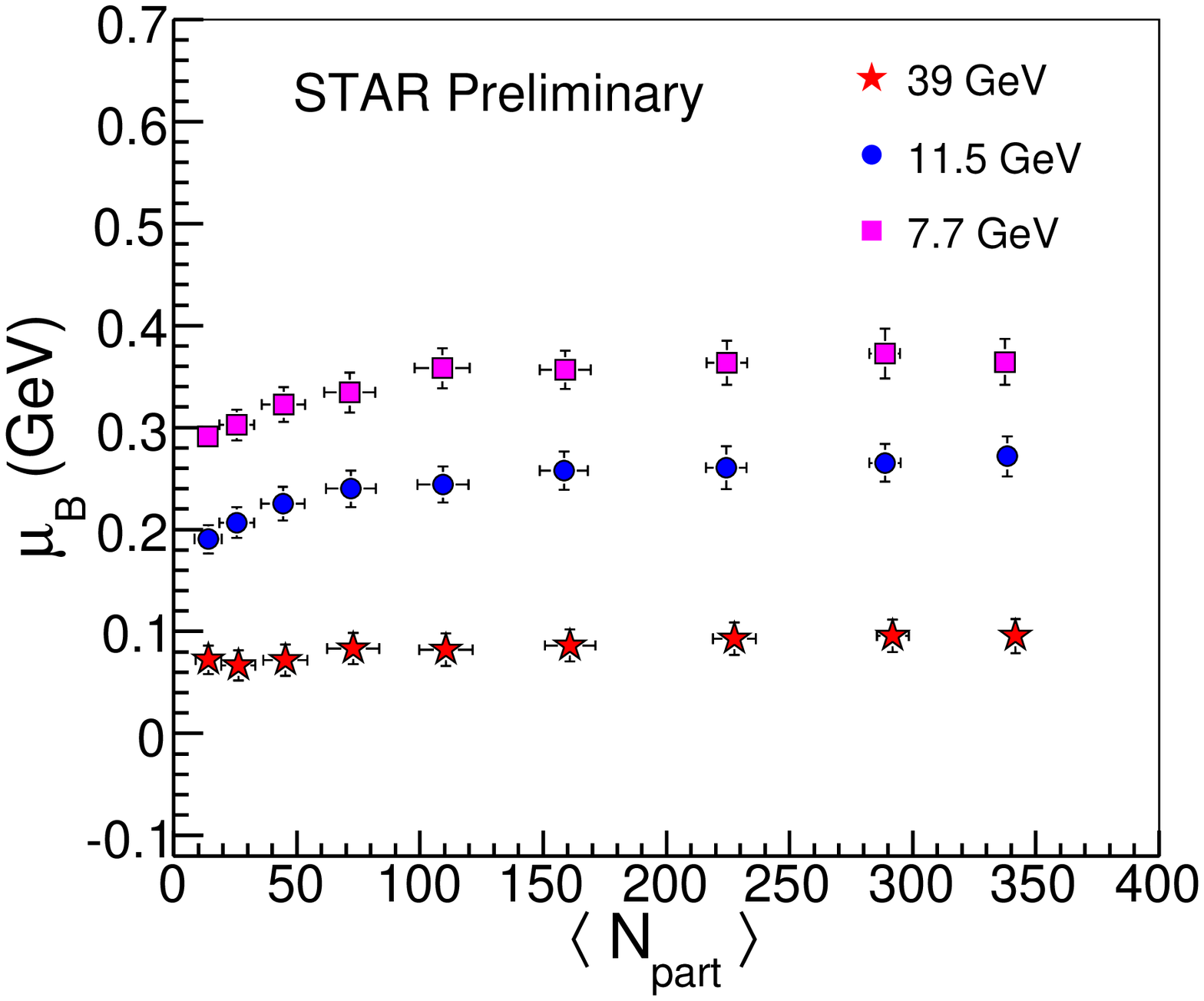}
\caption{Left: $T_{\rm{ch}}$ as function of $\langle N_{\rm{part}}
  \rangle$ for $\sqrt{s_{NN}}=$7.7, 11.5 and 39 GeV. Right: $\mu_{\rm{B}}$ as function of $\langle N_{\rm{part}}
  \rangle$ for $\sqrt{s_{NN}}=$7.7, 11.5 and 39 GeV.
Errors represent the statistical and systematic errors added in quadrature.
\label{npartfigs}}
\end{figure}

Figure~\ref{npartfigs} shows the centrality dependence of chemical
freeze-out temperature (left panel) and baryonic chemical potential
(right panel). The $T_{\rm{ch}}$ increases with increasing energy. It
also shows a slight increase as we go from peripheral to central
collisions for all energies. The baryonic chemical potential increases
with decreasing energy. This is because of large baryon stopping at
mid-rapidity at low energies. The $\mu_{\rm{B}}$ also shows a slight increase from
peripheral to central collisions for these energies.

Figure~\ref{fo_centdep} (left panel) shows the $T_{\rm{ch}}$ vs. $\mu_{\rm{B}}$
plot. Results are shown for Au+Au collisions at $\sqrt{s_{NN}}=$7.7,
11.5, 39 GeV and 200 GeV for different centralities. We see that the 200
GeV data do not show any centrality dependence for $T_{\rm{ch}}$ and
$\mu_{\rm{B}}$. The 39 GeV data show a slight dependence of freeze-out
parameters on centrality. However, when we go towards lower energies
(11.5 and 7.7 GeV), we observe a clear centrality dependence of the $T_{\rm{ch}}$ and
$\mu_{\rm{B}}$. This is 
the first observation of this behavior
by any experiment. It may
be noted that the freeze-out parameters shown here are obtained only
from the ratios involving $\pi, K, p$ and $\bar{p}$. 
It will be interesting to see the behavior of these parameters after
including
more ratios involving strange particles such as $\Lambda$ and $K^{0}_{s}$ in THERMUS fits.

\subsection{Kinetic freeze-out}
\begin{figure}
\includegraphics[scale=0.35]{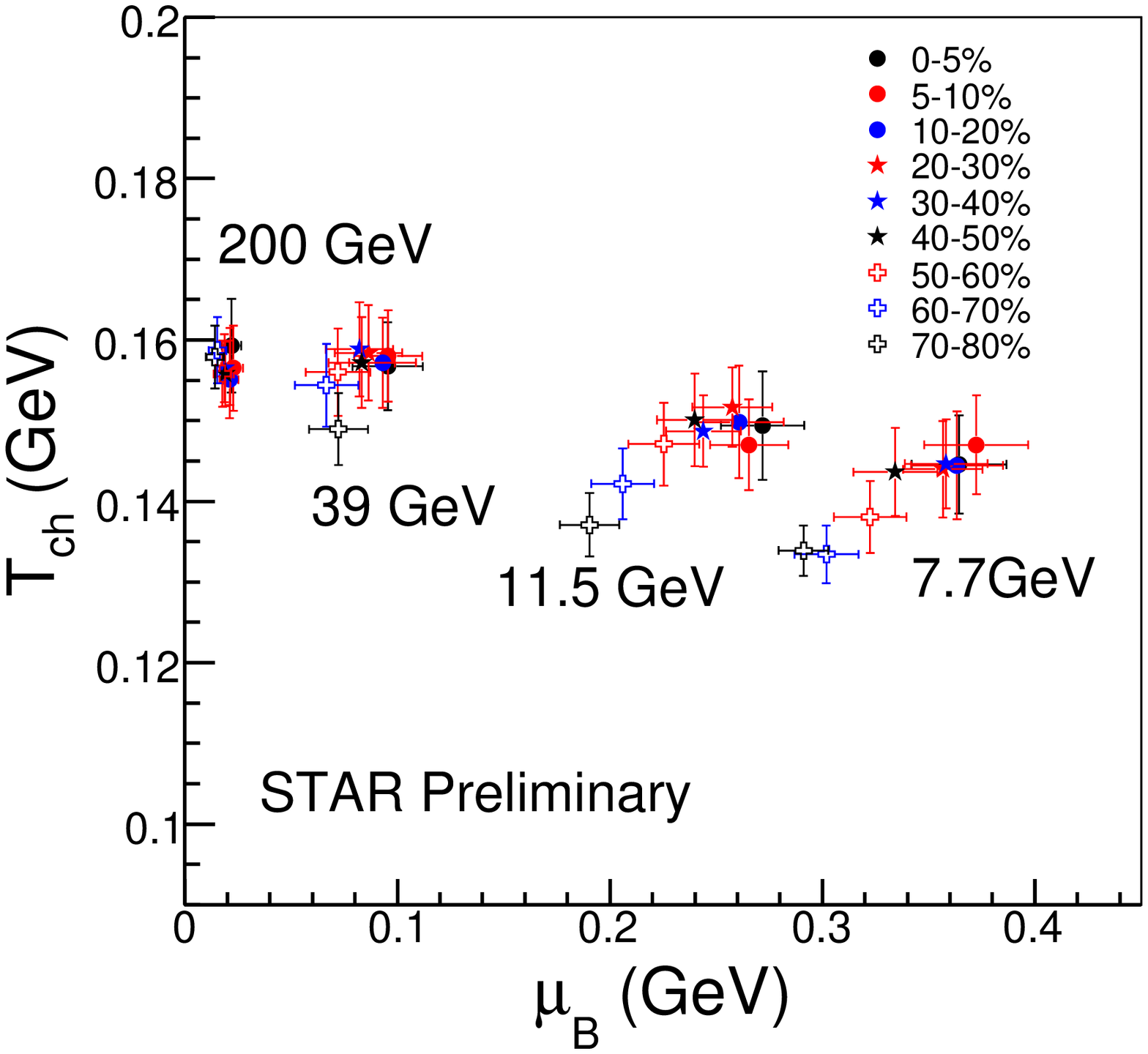}
\includegraphics[scale=0.35]{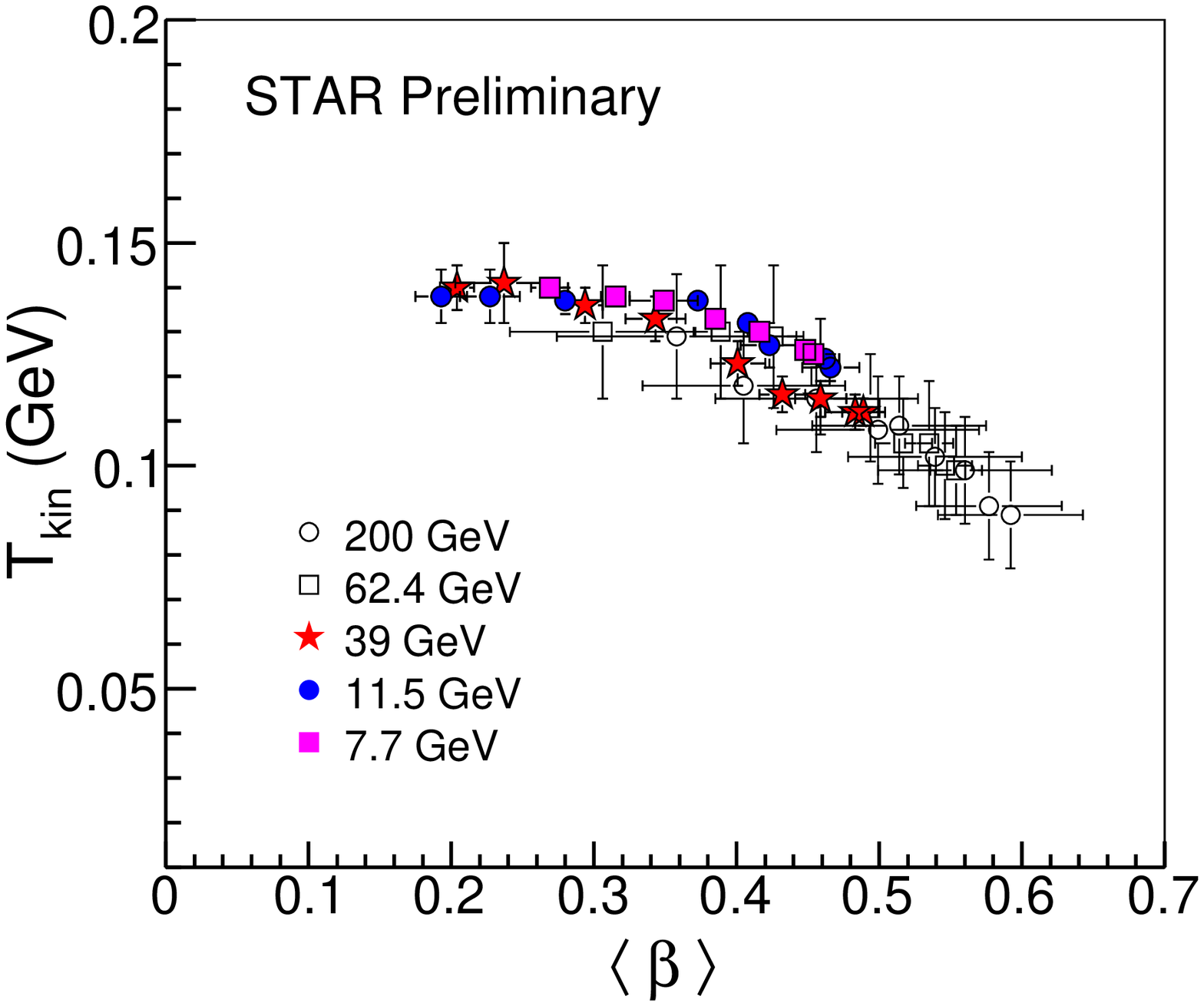}
\vspace{-0.5cm}
\caption{Left: Variation of $T_{\rm{ch}}$ with $\mu_{\rm{B}}$ for
  different centralities at $\sqrt{s_{NN}}=$7.7, 11.5, 39 and 200
  GeV. Right: Variation of $T_{\rm{kin}}$ with $\langle\beta \rangle$ for
  different centralities at $\sqrt{s_{NN}}=$7.7, 11.5, 39, 62.4 and
  200 GeV. The 62.4 and 200 GeV results are taken from the
  Ref.\cite{starpid}.
Errors represent the statistical and systematic errors added in quadrature.
\label{fo_centdep}}
\end{figure}

Kinetic freeze-out represents the scenario when the interactions among
the particles cease. After this time, the spectral shapes of the
particle species do not change. The kinetic freeze-out conditions are
obtained by fitting simultaneously the $\pi, K, p$ spectra with the
Blast-Wave model calculations. 
The BW model describes the spectral shapes assuming a locally
thermalized source with a common transverse flow velocity field.
The main fit parameters are the kinetic
freeze-out parameter $T_{\rm{kin}}$, the average flow velocity $\langle
\beta \rangle$, and the velocity profile $n$. 
The right panel of Fig.~\ref{fo_centdep} shows the variation of kinetic
freeze-out temperature as a function of average flow velocity.
Results are shown for different energies (7.7, 11.5, 39, 62.4 and 200 GeV) and for different
centralities. 
$T_{\rm{kin}}$ decreases with increase in energy. It also decreases as we go from peripheral to
central collisions.
The $\langle\beta \rangle$ increases with
increase in energy as well as collision centrality. The figure suggests that the higher
value of $T_{\rm{kin}}$ corresponds to the lower value of
$\langle\beta \rangle$ and vice-versa.

\section{Summary}
In summary, the centrality dependence of freeze-out parameters
(kinetic and chemical) are presented for the BES energies 7.7, 11.5
and 39 GeV.
The chemical freeze-out conditions are obtained using the particle
ratios. 
For the first time, a clear centrality dependence of freeze-out
parameters is observed at low energies. New measurements have
extended the $\mu_{\rm{B}}$ range covered by the RHIC data from 20-400
MeV in the QCD phase diagram. 
The kinetic freeze-out
conditions are obtained by using the particle $p_{T}$ spectra. The
kinetic freeze-out temperature decreases with increase in energy and
centrality. Average flow velocity increases with increase in energy
and centrality. 

We acknowledge the support from DOE and DAE-BRNS project sanction No. 2010/21/15-BRNS/2026.

\end{document}